\documentclass[twocolumn,aps,floatfix,showpacs]{revtex4}
\usepackage[dvips]{graphicx}
\begin{document}
\title{Optical conductivity of the Fr\"ohlich polaron}
\author{
A.S.~Mishchenko$^{1,2}$, N.~Nagaosa$^{1,3}$, N.V.~Prokof'ev$^{4,2}$,
A.~Sakamoto$^{3}$, and B.V.~Svistunov$^{4,2}$ }
\affiliation{
$^1$CREST, Japan Science and Technology Agency (JST), 
AIST Tsukuba Central 4,
Tsukuba 305-8562, Japan
\\ $^2$Russian Research Center ``Kurchatov Institute", 123182,
Moscow, Russia \\
$^3$CREST, Department of Applied Physics, The University of Tokyo, 7-3-1
Hongo, Bunkyo-ku, Tokyo 113, Japan \\ 
$^4$ Department of Physics,
University of Massachusetts, Amherst, MA 01003, USA }

\begin{abstract}
We present accurate results for optical conductivity of the  
three dimensional Fr\"ohlich polaron in all coupling regimes.
The systematic-error free diagrammatic quantum Monte Carlo method is employed
where the Feynman graphs for the momentum-momentum correlation function in 
imaginary time are summed up. The real-frequency optical conductivity
is obtained by the analytic continuation with stochastic optimization.
We compare numerical data with available perturbative and 
non-perturbative approaches to the optical conductivity 
and show that the picture of sharp resonances due to relaxed 
excited states in the strong coupling regime is ``washed out''
by large broadening of these states. As a result, the 
spectrum contains only a single-maximum broad peak with peculiar 
shape and a shoulder.         
\end{abstract}


\maketitle
 
Since the seminal work of Landau \cite{landau} the polaron problem, 
i.e. renormalization of quasiparticle properties due to
coupling to phonons, has been attracting considerable attention 
serving as a testing ground for non-perturbative methods.
The Hamiltonian for polarons consists of terms describing
free particle and phonon states \cite{polaron}
(Planck's constant and  electric charge are set to unity)
\begin{equation}
H_{0} \, = \, \sum_{\bf k} \, \varepsilon({\bf
k}) \, a^{\dag}_{\bf k} a^{ }_{\bf k} \; 
+
\sum_{\bf q} \, \omega_q \, b^{\dag}_{\bf q}  b^{ }_{\bf q} \; 
\label{H0}
\end{equation}
which interact through the particle-phonon coupling 
\begin{equation}
H_{\mbox{\scriptsize e-ph}} \, = \,
\sum_{{\bf k},{\bf q}} \,\left(   V({\bf q})\, 
 b^{\dag}_{\bf q}\, a^{\dag}_{{\bf k}-{\bf q}} a^{ }_{\bf k}
 + H.c.  \right) \,  \;.
\label{e-ph}
\end{equation}
In Eqs.~(\ref{H0}-\ref{e-ph}), $a^{ }_{\bf k}$ and $b^{ }_{\bf q}$
are the particle and phonon annihilation operators in momentum
space, respectively. The most popular models are the following two:
(i) Holstein lattice polaron (characterized by the 
tight-binding dispersion law 
$\varepsilon({\bf k})=2t\sum_{i=x,y,z} \cos (k_i)$ with hopping 
amplitude $t$, dispersionless optical phonons 
$\omega_{\bf q}=\omega_0$, and short-range interaction vertex 
$V_{\mbox{\scriptsize H}}({\bf q})=\gamma$),  
and (ii) the continuous Fr\"{o}hlich polaron (FP)
(characterized by $\varepsilon({\bf k})=k^2/2m$, 
$\omega_{\bf q}=\omega_0$, and long-range interaction
$V_{\mbox{\scriptsize F}}({\bf q})= 
i\omega_0 \big[ 2\sqrt{2}\pi\alpha \sqrt{m\omega_0} \big]^{1/2}/q$;
the convention is to measure energies in units of $\omega_0$,
and length scales in units of $1/\sqrt{m \omega_0}$).  
 
Although there are many reliable methods of calculating 
ground-state properties of polarons, the latter are 
hard to test directly in experiments. 
In contrast, the experimental data on optical conductivity (OC) spectra, 
$\sigma_{\beta \delta} (\omega )$, are readily available, 
but most theoretical methods can not well treat the excited states, 
and, as a result, are less accurate in predicting $\sigma (\omega) $. 
For decades, going back to early papers \cite{Fey62,Gur62,polaron}, 
the polaron OC (especially within the Fr\"{o}hlich model
\cite{Dev66,KED69,Dev72,Dev96})
constantly attracts attention of theoretical community. 

Even in the strong-coupling limit the results on OC are 
full of contradictions. 
In the variational adiabatic treatment of polarons at strong 
coupling \cite{Pekar51} one finds the so-called relaxed excited states (RES), 
i.e. quasi-stable states with the lattice distortion adjusted to the 
excited particle wavefunction. To contribute sharp peaks to OC spectra the 
decay rate of the lowest RES must be relatively small (otherwise the very
notion of RES becomes vague). The decay rate for RES  
with the energy $0.066\alpha^2$ \cite{DeRev72,Ev65,DevEv66} 
was calculated in 
the one-phonon approximation \cite{KED69} and a sharp peak in OC was 
predicted for $\alpha \ge 5$. 
Devreese, De Sitter, and  Goovaerts (DSG) \cite{Dev72} reproduced 
this result by the expansion of the impedance
$Z(\omega) \sim 1 / \sigma(\omega)$ within the  
Feynman-Hellwarth-Iddings-Platzman (FHIP) theory \cite{Fey62}.
However, the validity of the one-phonon approximation \cite{KED69} 
is in doubt for the strong coupling limit. 
Also, even within the {\it same FHIP approximation} \cite{Fey62}, 
the expansion of the inverse impedance 
does not show sharp features at 
energies predicted by the impedance expansion \cite{Dev72}. 
Finally, recent high-precision simulations of the Lehman spectral 
function \cite{MPSS}, $g(\omega )$, did not reveal any stable excited 
states for FP with the $\alpha^2$-dependence of energy - 
peaks with weak energy dependence on $\alpha$ were found instead.
It is worth mentioning, that the perturbation theory fails to reproduce
data on $g(\omega )$ even at very small $\alpha=0.05$. 
 
At present, the actual shape of OC for the Fr\"{o}hlich polaron 
is not known, and the crucial link connecting polaron theories to
experiments is missing. We are not aware of any systematic-error free 
method for calculating OC of continuous polarons. 
There are several non-perturbative approaches \cite{Bonca99,MPSS,Fehske00} 
dealing with the properties determined by excited states. 
However, the Lanczos diagonalization technique \cite{Fehske00} is limited 
to finite size systems, and the variational method \cite{Bonca99} encounters
difficulties for long-range interactions. 

In this Letter, we show that the problem may be solved by the 
Diagrammatic Monte Carlo method \cite{MPSS} based on direct summation 
of Feynman diagrams for the particle  momentum-momentum correlation 
function in imaginary time, $\langle k_{\beta}(\tau ) k_{\delta} (0) \rangle$, 
with subsequent analytic continuation to real frequencies by means 
of stochastic optimization. Our technique is generic and works for arbitrary
particle and phonon dispersion laws and interaction vertex. 
For the FP model we find that existing analytical methods severely 
underestimate multiphonon decay rates which broaden RES to the point     
that their contributions to OC overlap. Instead of sharp peaks
one obtains a picture of a broad single-maximum function with irregular shape
and a shoulder.   

\begin{figure}[th]
\hspace{-1.2cm}  \includegraphics{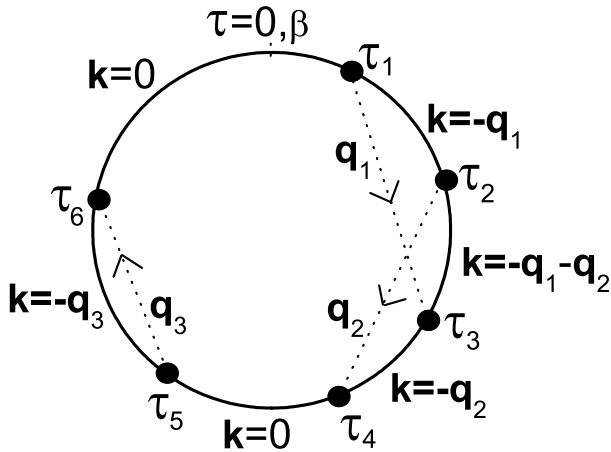} 
\caption{\label{fig:fig1}  A typical diagram for the polaron 
partition function. Solid (dotted) lines represent free particle 
(phonon) propagators. In general, there will be diagrams with all
particle propagators being accompanied by one or several phonon lines.} 
\end{figure}

The real part of $\sigma_{\beta \delta} (\omega)$ is straightforwardly 
related to $\langle k_{\beta}(\tau ) k_{\delta} (0) \rangle$
\begin{equation}
\sigma_{\beta \delta} (\omega) = (\pi / \omega) 
\langle 
k_{\beta} k_{\delta}
\rangle_{\omega} \; , 
\label{sigma}
\end{equation}
where (at zero temperature $\beta =1/T \to \infty $) 
\begin{equation}
\langle k_{\beta}(\tau) k_{\delta}(0) \rangle =
\int_{0}^{+\infty} d \omega \: e^{-\omega\tau}
\langle 
k_{\beta} k_{\delta}
\rangle_{\omega} \;.
\label{spectral}
\end{equation}
The solution of the integral Eq.~(\ref{spectral}) is obtained
by the stochastic optimization method of Ref.~\cite{MPSS,RP} which allows 
to handle both broad and infinitely sharp features in the spectrum on 
equal footing.  The $\langle k_{\beta}(\tau) k_{\delta}(0) \rangle$ 
correlator is diagonal in the particle momentum representation and
is readily  available (has a direct Monte Carlo estimator) from the 
statistics of Feynman diagrams for the partition function,
$Z = Tr \left\{ e^{-\beta H} \right\} \equiv 
Tr \left\{  e^{-\beta H_0} \hat{T}_{\tau }
\exp \left[ -\int_{0}^{\beta} H_{\mbox{\scriptsize e-ph}}(\tau) d \! \tau 
\right] \right\} $, (see Fig.~\ref{fig:fig1}). 
\begin{figure}[t]
\vspace*{-0.75cm}
\includegraphics{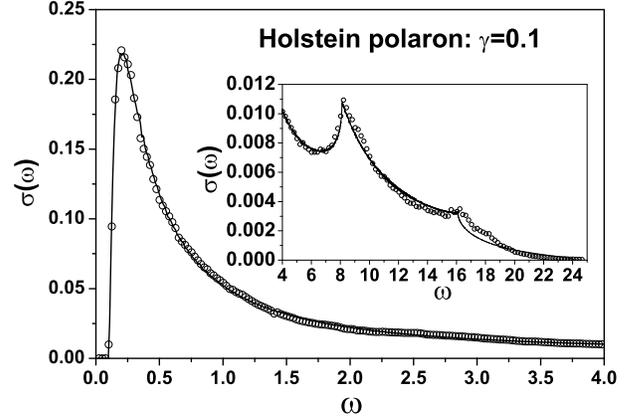} 
\caption{\label{fig:fig2} The low-energy part of OC for the Holstein
polaron (circles) compared to the perturbation theory result (line);
the high-energy tail is shown in the insert.
}
\end{figure}
\begin{figure}[tbp]
\includegraphics{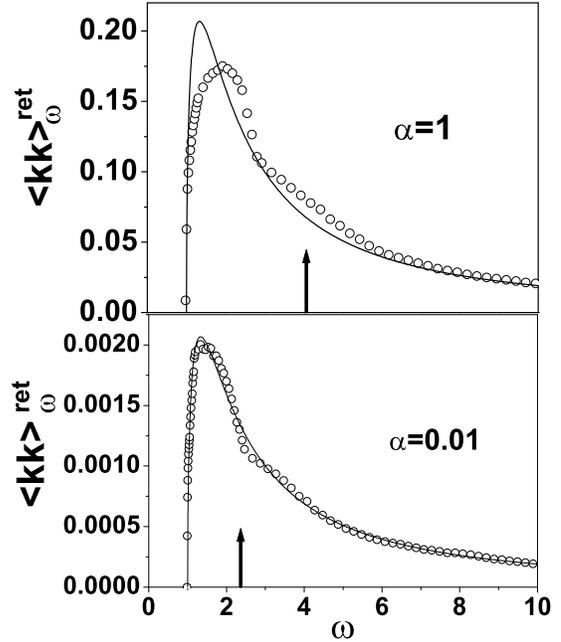}
\caption{\label{fig:fig3} Comparison between the calculated 
momentum-momentum correlator (circles) and the perturbation theory 
result (lines) for the Fr\"ohlich model. We indicate by arrows 
anomalies observed previously in the Lehman function.}
\end{figure}

Each graph generated by the
expansion of the $\tau$-ordered exponent contributes 
$\beta^{-1} \int_{0}^{\beta } d\tau' 
k_{\beta} (\tau' + \tau ) k_{\delta} (\tau' )$ to the correlator, 
where ${\bf  k} (\tau' )$ is the the step-wise function on the 
diagram circle. All calculations were done for finite but very large 
$\beta \omega > 100$, to ensure that exponentially small finite-temperature
corrections are negligible. 
The Metropolis-type sampling of 
diagrams in continuous momentum-time was performed using 
Diagrammatic Monte Carlo technique developed in 
Refs.~\cite{PS,MPSS}. We note, that formally our method works for
finite temperatures as well, but analytic continuation is notoriously
unstable at high temperatures ($\omega_0 T \sim 1$ in our case).
 
\begin{figure}[t]
\includegraphics{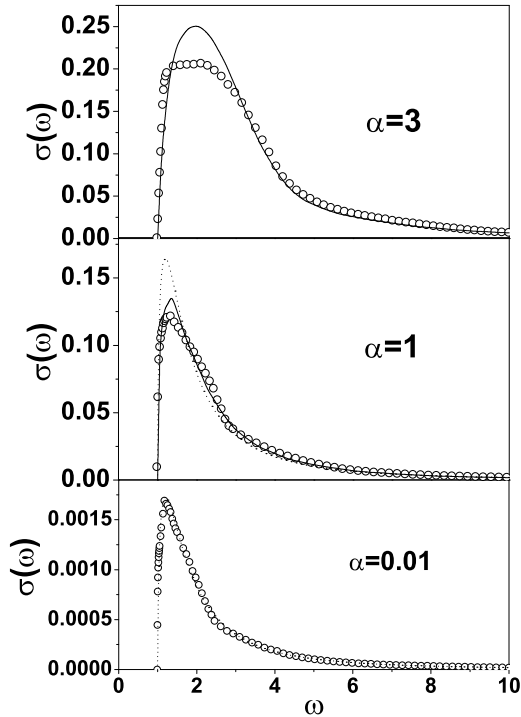} 
\caption{\label{fig:fig4}
OC data for the weak coupling regime (open circles)
compared to the second-order perturbation theory (dotted lines) 
and DSG calculations (solid lines).  }
\end{figure}

To test the method efficiency and reliability for the study of OC,
we simulated the three dimensional Holstein polaron in the weak 
coupling regime ($t=2$, $\omega_0=0.1$, 
$\gamma=0.1$) and compared our data in Fig.~\ref{fig:fig2}
to the perturbation theory result \cite{Mahan},
\mbox{
$\sigma_{\alpha \alpha}(\omega) = \sigma (\omega) = 
(\gamma^2/8\pi^2\omega)\int_{\mbox{\scriptsize BZ}}
d^3 \!q\, q_{\alpha}^2 \,\delta(\omega-\omega_0-\varepsilon({\bf q}))
$}, where the integral is taken over the Brillouin zone.
Not only the low-energy part is in perfect agreement with the theory, but
even sharp features in the tail due to van Hove singularities are
reproduced with remarkable (for analytic continuation procedures) 
accuracy.  

There is no such agreement with the perturbation theory for the
Fr\"ohlich polaron even at extremely small coupling constant 
$\alpha =0.01$ (see Fig.~\ref{fig:fig3}). This is due to 
the singular $1/q$-dependence of the coupling constant.
We note that positions 
of anomalies in $\sigma (\omega )$  and  $g(\omega )$ 
(see Ref.~\cite{MPSS}) spectra correlate.  
%
Apart from weak anomaly around $\omega \sim 3$, the OC spectrum is 
in reasonable overall agreement with the second-order perturbation 
theory result for $\alpha=0.01$ (see Fig.~\ref{fig:fig4}). 

For $\alpha=1$ our results deviate from the 2-nd order 
perturbation theory considerably; the  agreement with the DSG method
of impedance expansion \cite{Dev72} is much better. 
Since fourth-order corrections are very small in this parameter range
and do not lead to the DSG curve \cite{DeRev72}, we conclude that  
two-phonon terms are not sufficient to account for the discrepancy
and higher-order terms have to be taken into account even at $\alpha=1$.
For $\alpha=3$ the spectrum broadens considerably and its shape 
in the maximum starts deviating from the DSG result.

\begin{figure}[t]
\includegraphics{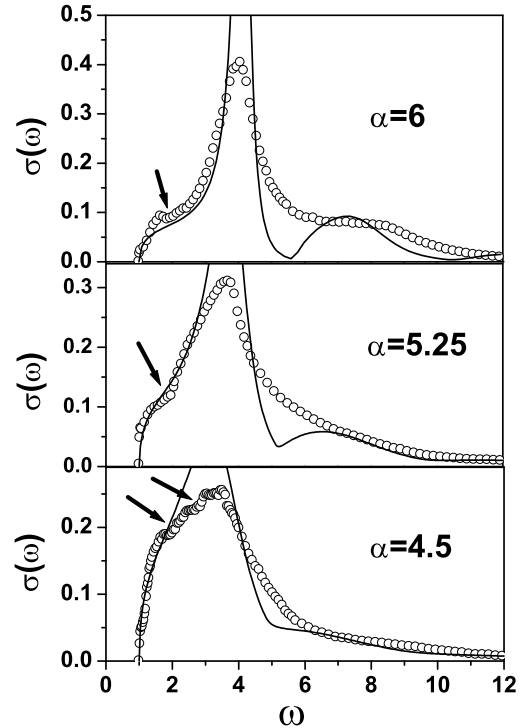} 
\caption{\label{fig:fig5} OC data for the 
intermediate coupling regim (open circles)
compared to the DSG approach (solid line).
Arrows point to the anomalies in absorption spectra arising at the  
two- and three-phonon thresholds. }
\end{figure}

\begin{figure}[t]
\includegraphics{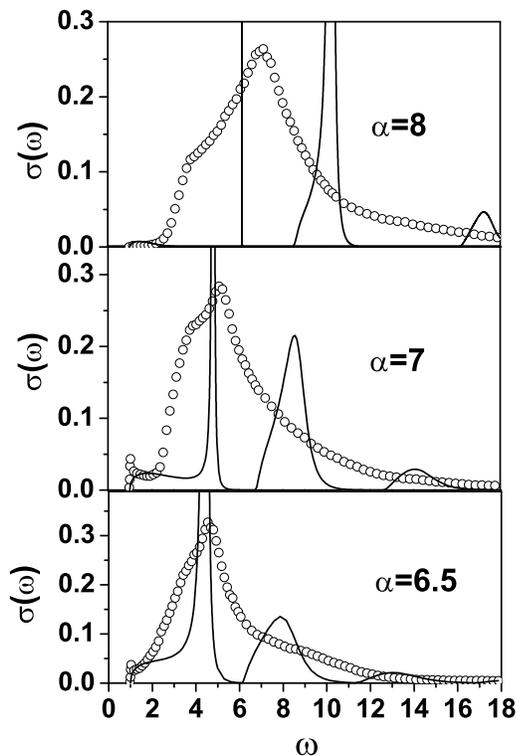} 
\caption{\label{fig:fig6} OC data for the strong coupling regime 
(open circles) compared to the DSG results (solid lines). 
}
\end{figure}
    
In the intermediate coupling regime $ 3 < \alpha < 7 $
we observe that the position of the spectrum maximum 
does follow the DSG prediction (and that of other papers advocating
the RES concept \cite{DevEv66,Ev65,KED69,DeRev72}) for the largest 
peak, and the low-energy behavior is also captured by the theory
reasonably well. However, the second peak 
at higher frequencies never actually fully develops, and, at best, 
its remnants are seen as a flat shoulder for $\alpha=6$.  
The other qualitative difference between our results and  DSG 
theory is significantly larger broadening of the dominant peak.
We stress, that broadening in our data is not an artifact of the 
analytic continuation method because we clearly see thresholds 
corresponding to two- and three-phonon emission/absorption processes 
(see Fig.~\ref{fig:fig5}). The conclusion we draw from this observation 
is that excited states are short-lived and their decay rates mediated by
multi-phonon processes are not accounted for in 
\cite{DevEv66,Ev65,KED69,DeRev72}. 
 
In the strong coupling regime, the broadening of excited states 
completely changes the overall shape of the spectrum 
(see Fig.~\ref{fig:fig6}). Instead of the double-peak structure
with extremely narrow central peak and a pseudogap,   
$\sigma (\omega )$ features a broad single-maximum
function away from the threshold. 
Even more surprisingly, the shoulder at the 
high-frequency end of the spectrum vanishes, and
another shoulder emerges at the low-frequency end (but away 
from the threshold at $\omega =\omega_0$). The central peak and the new
shoulder can be fitted (somewhat arbitrarily) as resulting from 
two strongly overlapping Gaussian peaks. In any event, 
the concept of quasi-stable excited states clearly has no place 
even at large $\alpha =8$ (the effective mass is 
already heavily renormalized at this point by a factor of thirty 
\cite{MPSS}).     
 
We conclude that physically attractive picture of stable RES, 
naturally emerging from the variational treatment of the strong coupling limit, 
may not be used to interpret OC spectra of Fr\"ohlich polarons.      
Decay rates of excited states are severely underestimated in
the one-phonon approximation \cite{KED69,DeRev72}. 
We may not rule out the possibility that at larger values
of $\alpha$ sharp peaks do appear in OC. However, this 
outcome would be rather academic because for $\alpha >15$ the 
effective mass renormalization is enormous, 
$m^*(\alpha =15 ) \sim 10^3 m$, and lattice effects enter the problem. 
In realistic models of polarons one also has to include into the picture
(i) several optical modes, (ii) interactions with acoustic phonons,
(iii) screening of the interaction potential, (iv) finite lattice constant,
etc. The simulation method developed in this Letter allows one to study  
all of the above mentioned effects.  Therefore the accurate analysis of
the experimental data is possible, which is left for future studies.
We note, that formally our method works for
finite temperatures as well, but analytic continuation is notoriously
unstable at high temperatures ($\omega_0 T \sim 1$ in our case).

We acknowledge fruitful and inspiring discussions with 
J. Devreese. This work was supported by  
RFBR  01-02-16508 and the National Science Foundation Grant No.\ 
DMR-0071767.


\begin{thebibliography}{99}

%
\bibitem{landau} L.D.\ Landau, Phys.\ Z.\ Sowjetunion {\bf 3} 664 (1933).
%
\bibitem{polaron} J.\ Appel, in {\it Solid State Physics},
              edited by H.\ Ehrenreich, F.\ Seitz, and D.\ Turnbull
              (Academic, New York, 1968), Vol. 21.
%
\bibitem{Fey62} R.\ Feynman, R.\ Hellwarth, C.\ Iddings, and P. Platzman,
                Phys.\ Rev.\ {\bf 127}, 1004 (1962)
%
\bibitem{Gur62} V.\ L.\ Gurevich, I.\ G.\ Lang, and Yu.\ A.\ Firsov, 
             Fiz.\ Tverd.\ Tela {\bf 4}, 1252 (1962)
            [Sov.\ Phys.\ - Solid State {\bf 4}, 918 (1962)]
%
\bibitem{Dev66} J.\ Devreese and R.\ Evrard, 
                Phys.\ Lett.\ {\bf 11}, 278 (1966).
%
\bibitem{KED69} E.\ Kartheuzer, R.\ Evrard, and J.\ Devreese,
                 Phys.\ Rev.\ Lett.\ {\bf 22}, 94 (1969).
%
\bibitem{Dev72} J.\ Devreese, J.\ De Sitter, and M.\ Goovaerts,
              Phys.\ Rev.\ B {\bf 5}, 2367 (1972).
%
\bibitem{Dev96} J.\ T.\ Devreese, in {\it Encyclopedia of Applied Physics},
               (New York: VCH Publishers, 1996) Vol.\ 14, pp.\ 383--413. 
%
\bibitem{Pekar51} S.\ I.\ Pekar, {\it Untersuchungen \"uber die
Elektronentheorie der Kristalle}, (Akademie Verlag, Berlin, 1954).
%
\bibitem{DeRev72} J.\ T.\ Devreese, in {\it Polarons in Ionic crystals and 
                 Polar Semiconductors} (North Holland, Amsterdam, 1972). 
%
\bibitem{Ev65} R.\ Evrard, Phys.\ Lett.\ {\bf 14}, 295 (1965).
%
\bibitem{DevEv66} J.\ Devreese and R.\ Evrard, Phys.\ Lett.\ {\bf 11}, 
                  278 (1966).
%
\bibitem{MPSS} A.\ S.\ Mishchenko, N.\ V.\ Prokof'ev, 
            A.\ Sakamoto, and B.\ V.\ Svistunov, 
            Phys.\ Rev.\ B {\bf 62}, 6317 (2000).
%
\bibitem{Bonca99} J.\ Bonca, S.\ A.\ Trugman, and I.\ Batistic,
            Phys.\ Rev.\ B {\bf 60}, 1633 (1999). 
%
\bibitem{Fehske00} H.\ Fehske, J.\ Loos, and G.\ Wellein,
             Phys.\ Rev.\ B {\bf 61}, 8016 (2000). 
%
\bibitem{RP} A.\ S.\ Mishchenko, N.\ Nagaosa, N.\ V.\ Prokof'ev, A.\ Sakamoto,
             and B.\ V.\ Svistunov, 
             Phys.\ Rev.\ B {\bf 66}, 020301(R) (2002).
%
\bibitem{PS} N.\ V.\ Prokof'ev and B.\ V.\ Svistunov,
             Phys.\ Rev.\ Lett.\ {\bf 81}, 2514 (1998).
%
\bibitem{Mahan} G.\ D.\ Mahan, {\it Many particle physics} (Plenum Press,
New York, 1990). 
%
\end{thebibliography}
\end{document}